# Building an Ontology for the Domain of Plant Science using Protégé

Sara Hosseinzadeh Kassani
Department of Computer Science
University of Saskatchewan
sara.kassani@mail.usask.ca

Peyman Hosseinzadeh Kassani
Department Biomedical Engineering
Tulane University
peymanhk@tulane.edu

*Abstract*—Due to the rapid development of technology, large amounts of heterogeneous data generated every day. Biological data is also growing in terms of the quantity and quality of data considerably. Despite of the attempts for building uniform platform to handle data management in Plant Science, researchers are facing the challenge of not only accessing and integrating data stored in heterogeneous data sources, but also representing the implicit and explicit domain knowledge based on the available plant genomic and phenomic data. Ontologies provide a framework for describing the structures and vocabularies to support semantics of information and facilitate automated reasoning and knowledge discovery. In this paper, we focus on building an ontology for Arabidopsis Thaliana in Plant Science domain. The aim of this study is to provide a conceptual model of Arabidopsis Thaliana as reference plant for botany and other plant sciences, including concepts and their relationships.

*Keywords— plant ontology, semantic web, ontology engineering, knowledge modeling*

## I. INTRODUCTION

The process of organization, administration and governance of very large data sets is known as big data management. Many science domains can benefit from more efficient storing and analyzing big data to derive meaningful [1][2]. The complex and highly interrelated data accumulation have been incontrovertibly changing the traditional management strategies. Most traditional data management strategies do not sufficiently consider dynamic scalability and availability which is highly demanded by Big Data [3][4]. In the absence of appropriate big data management strategies support, the decision-making process has become cumbersome and ineffective task [2]. As the amount of data grows, the need for a scalable infrastructure for extracting semantics becomes stronger and several challenges are experienced by researchers [5]. For example, it is difficult for scientists to find an efficient method to analyze various scientific data formats in a complete and accurate manner. Building an integrated framework with the help of ontology can reduce these difficulties in a specific domain of science. According to Tom Gruber: "An ontology is a specification of a conceptualization". The domain ontology is a formal representation of interrelated concepts within a specific domain and describing the relationships between those concepts [2], which aims at supporting shared and explicit description of concepts to aid in knowledge acquisition and reasoning. Ontology offers a unified model that would significantly leverage the power of the knowledge representation from vast amounts of data collected in plant sciences [6]. Building an ontology for capturing plant anatomical and morphological entities and also development stages for plant structures such as leaf or flower development stage help researchers to map the relationship between plant phenotype and genotype and represent knowledge [7].

## II. BACKGROUND

The World Wide Web (WWW) is a large-scale level of shared information, which is related by Hypertext links on the web. Hypertext links between web pages are a powerful tool that allow users to explore through highly heterogenous data in a wide variety of domains of science. A user can click on a Hypertext link and be referred to another link web page. Originally, this interlinked information on the web is comprehensible by humans, but it is almost impossible for machines to interpret the information. The lack of machine-readable information over the web motivated an ongoing effort to provide machine interoperability tools and technology based on ontology inference [8]. Semantic Web technologies empower researchers to specify logical and well-defined semantics, and thus enable computers to understand information in an automated manner. Tim Berners-Lee, the inventor of the World Wide Web proposed the term "Semantic Web", an information integrated infrastructure in which data can be processed by inference-enable computers[9] based on its meaning [10]. The Semantic Web is an extension of the traditional Web technology, annotating each web resource with machine-interpretable semantic metadata to provide a more expressive language to better reason about data items and the relations existing between them. Such semantic metadata could be utilized to facilitate the integration and reasoning of data derived from multiple sources [11].

The four main technologies of Semantic Web are URIs, RDF, OWL, and SPARQL. URIs are used to uniquely label entities in Linked Data. The Resource Description Framework (RDF), a World Wide Web Consortium (W3C) recommended standard, is an unordered, node and edge labeled graph-based data model which provides a basic representation of knowledge to describe entities, classes and their properties and also relationship to other data entities [12]. RDF expressions, which is called a triplet, composed of subject-predicate-object expressions for publishing data [11] as expressed visually in Figure 1. OWL (Web Ontology Language) is a description logic [7] language for defining ontologies. SPARQL is a pattern-matching query language for querying RDF graphs and extract information with high accuracy [13]. RDF and OWL are expressive enough for data integration and annotating data items with rich semantics.

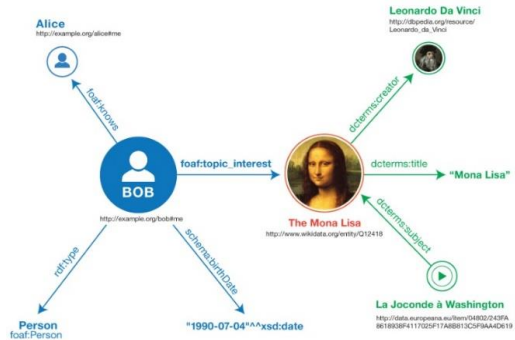

Figure 1. RDF triple graph

(Source: https://www.w3.org)

Therefore, development of an ontological infrastructure to support the semantic web has received extensive attention particularly with a rise of bioinformatics databases in life science. Ontology plays an important role to leverage automated reasoning to extract meaningful biological and biomedical information [14].

### III. ONTOLOGY IN PLANT SCIENCE

The term Ontology is originated from philosophy where it refers to the nature of existence. Particularly, ontology was used to provide a semantic framework for representing knowledge using ontology representation languages [15]. Currently, several ontology representation languages have been proposed including RDF, RDFS and OWL to capture the semantics of the domain of study. The bio-ontology has been emerged for enhancing the interoperability within biological knowledge with best practices on ontology development. The most cited bio-ontology is the Gene Ontology (GO), which is a tool for the unification of biology developed by Gene Ontology Consortium in 2000, present more than 30000 species-independent control vocabularies for describing gene products including plants (Gene Ontology Consortium, 2009). Plant Ontology (PO) focused on develop and share unambiguous vocabularies for plant anatomy and morphology. PO consist of two sub-categories: the plant structure ontology and the growth and developmental stages ontology. By defining classes of entities, logical relations, properties, constraints and range axioms, botany and plant science researchers are able to understand, share and reuse knowledge in a machine or computer interpretable content, enabling them to detect and reason biologically common concepts in heterogenous data sets [7].

The Plant Science Ontology main goal is to design and develop a semantic framework in order to support computerized reasoning. With the help of ontologies, scientists are able to employ the PO or GO as a general reference to semantically link large amounts of plant phenotype and genotype data together. However, knowledge engineering requires an extensive knowledge of different domains such as biology, engineering and also standard ontology languages. In this context, the Semantic Web also advocate efforts of developing knowledge management systems for capturing and extracting biological knowledge from highly distributed data resources [7].

### IV. TOOLS AND IMPLEMENTATION

There is no specific method for modeling and building a domain ontology, and the majority of the best methodologies for developing an ontology are depend on purpose of research. For building an ontology, researchers should consider three features. First, identifying the domain and scope of the ontology. Second choosing the language and logic to construct the ontology. Finally, identifying key concepts of resources (nouns) and relationships (verbs), and domain and range axioms [16].

We study the Arabidopsis Thaliana as reference plant for building our ontology. Arabidopsis Thaliana is the best-investigated of flowering plant species, belonging to the Brassicaceae family. It has been chosen as one of the most widely used model plant organism for studies in plant research in areas such as developmental and molecular genetics analysis, population genetics and genomics for many years. Arabidopsis Thaliana has five chromosomes, a fully sequenced genome. Its significant properties such as short regeneration time, simple growth requirements make it desirable for model plant studies. Therefore, studying biological processes in this species is important for gaining information of plant science and for utilizing of this knowledge to other relevant plants species [17].

Ontology tools facilitated constructing ontologies during the ontology development process. We have used the Protégé as the principal ontology authoring tool in our ontology-based application. Protégé is an IDE developed at Stanford University by Stanford Medical Informatics team. It is a free, open source ontology platform, which enable users to create and populate ontology and formal knowledge-based applications more straightforward. There exist many plug-ins for Protégé offering a number of powerful features [18]. We used the OWLViz plug-in to visualize Protégé ontologies. OWLViz is a powerful and highly configurable extension providing a graphical representation of the sematic relationships for helping users to visualize classes in an OWL ontology. OWLViz output created knowledge-based graph of the classes to different formats such as JPEG and PNG [19].

The process of building ontology for Arabidopsis Thaliana with the top-down approach has been described in detail in this section. This ontology acts as a basis for researchers to conduct query and reason in a knowledge based environment. All OWL classes inherit from a single root class called owl:Thing. The class owl:Thing represent the concept of any user-defined class or individuals in order to facilitate reasoning, as illustrated in Figure 2.

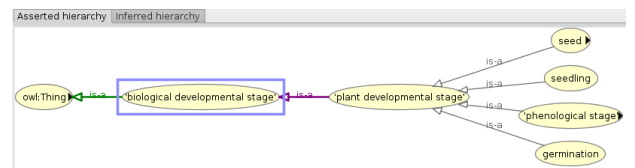

Figure 2 Calss Tree of Arabidopsis Thaliana in Protege

Owl:Thing is developed by W3C located at http://www.w3.org/2002/07/owl as part of OWL vocabulary and is equivalent to rdfs:Resource. The most basic part of the ontology for Arabidopsis Thaliana ontology is super classes such as biological developmental stage, biological process, and biochemical process as shown in Figure 3. Each super class consists of sub classes which arranged in an inheritance hierarchy. The second level in ontology are subclasses which provide more refined and detailed information about superclass, such as germination, life span, seed and seedling.

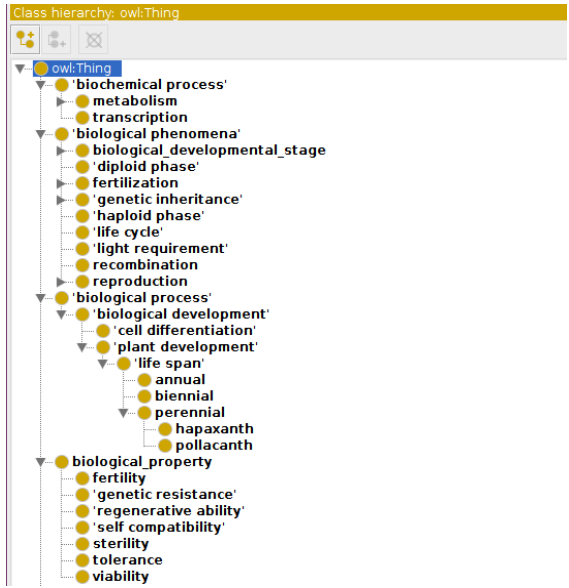

Figure 3 Calss Hierarchy of Arabidopsis Thaliana in Protege

OWL properties express association between two entities of a domain. Different types of properties are used to link between concepts. Figure 4 shows the list of declared properties of Arabidopsis Thaliana ontology in Protégé. Some of the relations have been used in this ontology are <growsIn>, <hasPart>, and <hasVariant>.

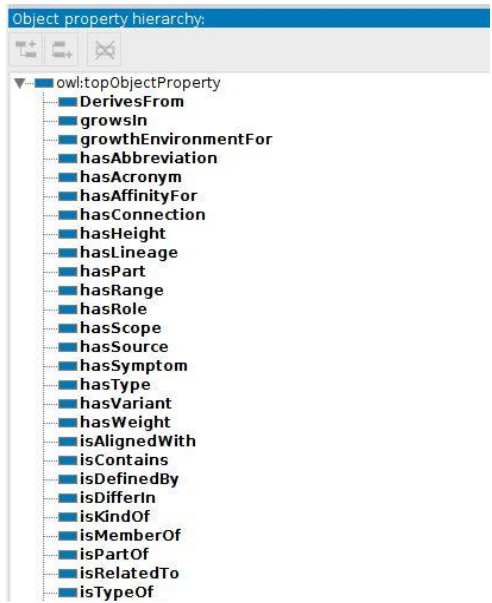

Figure 4 Object Properties of Arabidopsis Thaliana

Domain and range constraints of the properties aid to precisely describe representation of knowledge. Domain indicates the type of individuals that can be the subject and range specifies the type of individuals that can be the object within the RDF triple. The individuals are the members or instances of a class with certain constraints. See Figure 5.

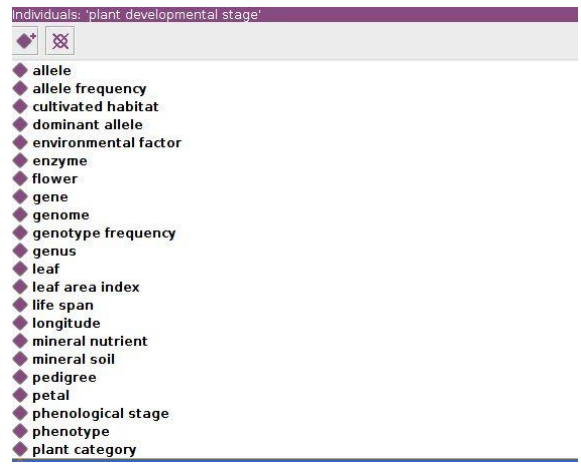

Figure 5 Individuals of Arabidopsis Thaliana Ontology

After preparing the ontology, we have used Apache Jena Fuseki server, a HTTP-based query engine, for executing the SPARQL queries over the web of linked data and extract triple information on the web page. Apache Jena Fuseki server is a SPARQL server providing hosts for persistent storage or in-memory storage of the datasets and aimed toward supporting developers with building practical semantic web applications.

The class biological property, for example, has several subclasses. Some subclasses of biological property that are used to describe the class are such as genetic resistance, regenerative ability, seed compatibility, tolerance and viability. A wide variety of object properties are used to describe relationships between classes and subclasses. The object property growsIn describe the plant specific requirement for growth and development. The object property maxHeight indicates the plant sample height growth pattern. The ontology of Arabidopsis Thaliana provides detailed information pertaining to the morphology and developmental stages.

V. ONTOLOGY EVALUATION

After building the Ontology, we should assess the consistency and the quality of the ontology using Protégé reasoner. Ontology evaluation and validation ensure the avoidance of excessiveness concepts, terminological ambiguity, incompatible subclass relationships. Ontologies needs to be validated to conform a standardized OWL profiles, the expressivity of ontology, as well as the consistency of structure described in the model with expected semantics to support the exchange of information efficiently [20]. Further consultation with domain experts is needed to examine the scope of the proposed ontology. To evaluate the overall performance of the implementation and ontology syntax, we have used OntoCheck plugin to evaluate the consistency, conciseness. and correctness of ontology automatically.

OntoCheck is an open source plug-in developed at the University of Freiburg, and it is currently one of the W3C OWL official validating tool [21].

## VI. CONCLUSION AND FUTURE WORK

Building an ontology for plant science domain to determine concepts, properties or instances provide the botany and plant scientists the ability to efficiently inference and extract valuable knowledge over the data. In this study, the ontology editor Protégé and Apache Jena Fuseki server was employed to construct an ontology for growth and developmental stages of Arabidopsis Thaliana in Plant Science domain. The proposed work is expected to satisfy the requirements of formal representations of knowledge from semantics-enriched information. In future, we intend to combine semantic data mining and semantic data repositories expressed in RDF/(S) and OWL to discover relevant patterns from a biological setting, and to build data integration implementations based on further biological use cases.